\documentclass[letterpaper]{aa} 
\usepackage{graphicx}
\usepackage{txfonts}
\begin{document}
\title{A Massive Planet to the Young Disc Star HD 81040
\thanks{Based on observations made at the Observatoire de Haute-Provence (French CNRS)
        and at the W. M Keck Observatory, which is operated as a scientific partnership
        among the California Institute of Technology, the University of California and
        the National Aeronautics and Space Administration. The Observatory was made
        possible by the generous financial support from the W. M. Keck Foundation.}}


   \author{A. Sozzetti\inst{1,2}\and S. Udry\inst{3}  \and S. Zucker\inst{4,5} 
           \and G. Torres\inst{1} \and J. L. Beuzit\inst{6} 
           \and D. W. Latham\inst{1} \and M. Mayor\inst{3}
           \and T. Mazeh\inst{4} \and D. Naef\inst{7}  \and C. Perrier\inst{6}
           \and D. Queloz\inst{3}\and J.-P. Sivan\inst{8}}

   \offprints{A. Sozzetti, \\\email{asozzett@cfa.harvard.edu}}

   \institute{Harvard-Smithsonian Center for Astrophysics, 60 Garden Street,
              Cambridge, MA 02138, USA
\and INAF - Osservatorio Astronomico di Torino, 10025 Pino Torinese, Italy
\and Observatoire de Gen\`eve, 51 Ch. de Maillettes, 1290 Sauveny, Switzerland
\and School of Physics and Astronomy, Raymond and Beverly
     Sackler Faculty of Exact Sciences, Tel Aviv University, Tel Aviv 69978, Israel 
\and Department of Geophysics and Planetary Sciences, Beverly and Raymond Sackler 
     Faculty of Exact Sciences, Tel Aviv University, Tel Aviv 69678, Israel
\and Laboratoire d'Astrophysique de Grenoble, Universit\'e J. Fourier, BP 53,
     38041 Grenoble, France
\and European Southern Observatory, 3107 Alonso de Cordova, Casilla 19001, 
     Santiago 19, Chile
\and Observatoire de Haute-Provence, 04870 St-Michel l'Observatoire, France}

   \date{Received ......; accepted ......}

   \abstract{
We report the discovery of a massive planetary companion orbiting the young disc 
star HD 81040. Based on five years of precise radial-velocity measurements with
the HIRES and ELODIE spectrographs, we derive a spectroscopic orbit with a
period $P =1001.0$ days and eccentricity $e = 0.53$. The inferred minimum mass
for the companion of $m_2\sin i = 6.86$ M$_\mathrm{Jup}$ places it in the high-mass
tail of the extrasolar planet mass distribution. The radial-velocity residuals
exhibit a scatter significantly larger than the typical internal measurement
precision of the instruments. Based on an analysis of the Ca II H and K
line cores, this is interpreted as an activity-induced phenomenon.
However, we find no evidence for the period and magnitude of
the radial-velocity variations to be caused by stellar surface activity.
The observed orbital motion of HD 81040 is thus best explained with the presence
of a massive giant planet companion.
   \keywords{planetary systems -- stars: individual (HD 81040) -- stars: activity -- 
             stars: abundances -- techniques: radial velocities -- 
             techniques: spectroscopic}
   }

   \maketitle
%

\section{Introduction}

Except for one experiment (Cochran et al. 2002), radial-velocity
surveys for planets orbiting nearby, solar-type stars
with a targeted precision of  $1-5$ m s$^{-1}$ (e.g., Mayor et al. 2004;
Santos et al. 2004; Butler et al. 2004; Marcy et al. 2005a, 2005b) 
have systematically attempted to exclude from their samples
chromospherically active stars. This is because changes in the 
visibility of active regions and variations of the 
stellar absorption line profiles can possibly cause significant
deterioration in the achievable radial-velocity precision.
However, for about a dozen planet hosts significant activity levels
have not prevented radial-velocity surveys
from detecting Keplerian signals due to orbiting giant planets.
We report in this paper on radial-velocity measurements of the
young disc star HD 81040 (HIP 46076, BD+20\degr2314). These
observations reveal the presence of a massive planet 
candidate orbiting the star, 
with a minimum mass of 7.61 $M_\mathrm{Jup}$ (where
$M_\mathrm{Jup}$ is the mass of Jupiter).

The variable velocity of HD 81040 was first detected
by the G-Dwarf Planet Search (Latham 2000), a program
designed to conduct a first reconnaissance for giant planets
orbiting a sample of nearly 1000 nearby G dwarfs. This
survey employed the HIRES spectrograph (Vogt et al. 1994)
on the 10-m Keck 1 telescope at the W. M. Keck Observatory
(Hawaii). Follow-up observations were carried out within the
context of the ELODIE Planet Search Survey (Mayor \& Queloz
1995; Perrier et al. 2003), which uses
the ELODIE fiber-fed echelle spectrograph (Baranne et al. 1996)
on the 1.93-m telescope at the Observatoire de Haute-Provence
(CNRS, France).

Radial velocities with ELODIE are obtained by cross-correlating
the observed spectra with a numerical template. The ``simultaneous
thorium-argon technique'' with dual fibers (Baranne et al. 1996)
allows for the monitoring and correction of instrumental drifts.
The precision achieved with this instrument on bright, inactive
stars is $\sim 6.5$ m s$^{-1}$ (Perrier et al. 2003). 
The HIRES instrumental profile and
drifts are monitored using an iodine absorption cell (Marcy \&
Butler 1992). In order to derive radial velocities, the data
reduction procedure involves the modelling of the temporal and
spatial variations of the instrumental profile of the spectrograph
(Valenti et al. 1995), and is conceptually similar to that
described by Butler et al. (1996). Internal errors, computed from
the scatter of the velocities from the echelle orders containing
I$_2$ lines, are typically 8-10 m s$^{-1}$ for inactive,
solar-type stars, for the relatively short exposures adopted for 
the G-dwarf Planet Search.

The collaboration between our two teams has so far resulted in
three important discoveries: the determination of the spectroscopic
orbit of the first extrasolar planet transiting the disc of its
parent star (HD 209458b; Mazeh et al. 2000), the detection
of HD 80606b, the planet with the largest eccentricity
known to-date ($e = 0.927$; Naef et al. 2001a), and the identification
of the first giant planet in a stellar triple system
(HD 178911Bb; Zucker et al. 2002).

The Keck/HIRES observations of HD 81040 began in April
1999 and have a time baseline of $\sim 8.5$ months. The ELODIE
follow-up observations span $\sim 3.3$ years,
starting in February 2002. We present in Sect.~2 the
radial-velocity data and the orbital solution derived
utilizing the combined datasets. The stellar
characteristics of HD 81040 are considered in Sect.~3.
Finally, Sect.~4 is devoted to a summary and
discussion of our findings.

\section{Radial-Velocity Data and Orbital Solution}

\begin{table}
\caption{Radial velocity measurements of HD 81040.
The HIRES (H) velocities are shifted to the ELODIE (E) zero point.}
\label{tab1}
\centering
\begin{tabular}{l c c c}
\hline\hline
BJD - 2,400,000 & Radial Velocity & $\sigma_\mathrm{RV}$ & Instrument \\
& (km s$^{-1}$) & (km s$^{-1}$) & \\
\hline
51291.36542  & 49.319  &  0.012  &  H \\
51293.37376  & 49.312  &  0.014  &  H  \\
51545.63921  & 49.088  &  0.009  &  H \\
52308.63310  & 49.399  &  0.012  &  E \\
52356.41490  & 49.412  &  0.021  &  E \\
52359.45970  & 49.363  &  0.012  &  E \\
52360.42310  & 49.391  &  0.011  &  E \\
52615.71400  & 49.119  &  0.011  &  E \\
52649.66360  & 49.126  &  0.011  &  E \\
52649.67630  & 49.129  &  0.013  &  E \\
52719.40970  & 49.100  &  0.008  &  E \\
52722.49520  & 49.118  &  0.012  &  E \\
52993.63610  & 49.204  &  0.012  &  E \\
53034.56760  & 49.293  &  0.015  &  E \\
53094.36570  & 49.308  &  0.010  &  E \\
53101.39950  & 49.285  &  0.010  &  E \\
53359.69070  & 49.389  &  0.016  &  E \\
53361.66000  & 49.412  &  0.021  &  E \\
53421.52680  & 49.459  &  0.019  &  E \\
53428.48940  & 49.462  &  0.026  &  E \\
53461.36050  & 49.414  &  0.016  &  E \\
53463.41260  & 49.458  &  0.020  &  E \\
53464.39850  & 49.408  &  0.011  &  E \\
53486.35380  & 49.359  &  0.011  &  E \\
53491.32410  & 49.348  &  0.010  &  E \\
53518.35250  & 49.234  &  0.013  &  E \\
\hline
\end{tabular}
\end{table}

\begin{table}
\begin{minipage}[t]{\columnwidth}
\caption{Best-fit orbital solution for HD 81040, derived
minimum companion mass, and properties of the post-fit
residuals.}
\label{tab2}
\centering
\renewcommand{\footnoterule}{}  
\begin{tabular}{lc}
\hline\hline
Parameter & Value \\
\hline
$P$ (days)     &  $1001.7\pm 7.0$ \\
$T$ (BJD)  & $2452504.0\pm 12.0$ \\
$e$      &  $0.526\pm 0.042$ \\
$\gamma$ (km s$^{-1}$)     & $+49.2535\pm 0.0063$  \\
$\omega$ (\degr)     & $81.3\pm 7.2$  \\
$K_1$ (m s$^{-1}$)     &  $168\pm 9$ \\
$\Delta RV_\mathrm{H-E}$ (km s$^{-1}$)     & $+0.057\pm 0.019$  \\
$a_1\sin i$ ($\times 10^6$ km)     &  $1.96\pm 0.19$ \\
$f_1(m)$ ($\times 10^{-7}$ $M_\odot$)     & $3.0\pm 0.9$  \\
$m_2\sin i$ ($M_\mathrm{Jup}$)     &  $6.86\pm 0.71$ \\
$a$ (AU)     &  1.94 \\
$N$     &  23(E) + 3(H) \\
$<\sigma_\mathrm{RV}>$ (m s$^{-1}$)     &  13.7 \\
$\sigma_{O-C}$ (m s$^{-1}$)     &  26.0 (E: 25.0, H: 6.0) \\
$\chi^2_\nu$    &   3.26 \\
$Pr(\chi^2)$  &  $2\times 10^{-9}$   \\

\hline
\end{tabular}
\end{minipage}
\end{table}

\begin{figure}
   \centering
   \includegraphics[width=1.\columnwidth]{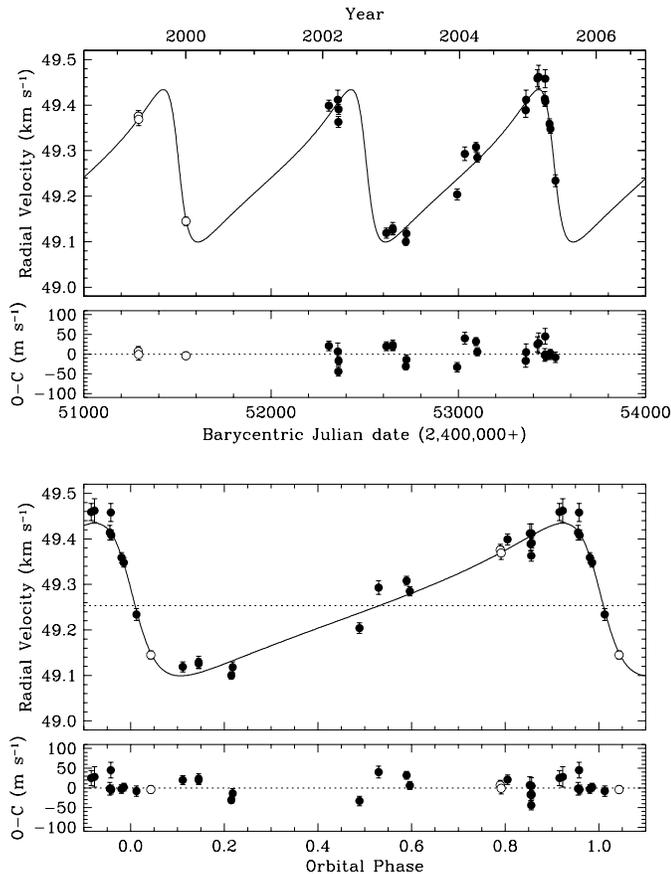}
\caption{Top: Radial-velocity measurements as a
function of time for HD 81040. Open circles identify
Keck/HIRES velocities, while filled black circles
correspond to the ELODIE dataset. Bottom: radial velocities
as a function of orbital phase. The two sub-panels
show the post-fit residuals in both cases.}
         \label{fig1}
   \end{figure}

The 26 radial velocity measurements for HD 81040 are listed in
Table~\ref{tab1}. In order to bring the HIRES data to the ELODIE
system, they have been initially shifted by an amount corresponding to the
$\gamma$ velocity measured by ELODIE, $\Delta RV = 49.24$
km s$^{-1}$. A residual velocity offset $\Delta RV_\mathrm{H-E}$
between the two is included as an additional free parameter in the
best-fit orbital solution presented in Table~\ref{tab2}. 
Formal uncertainties in the parameters are obtained by re-scaling 
by the reduced $\chi^2$ of the solution. 
We show in the upper and lower panel of Fig.~\ref{fig1} the
radial-velocity measurements as a function of time and orbital
phase, respectively. Post-fit velocity residuals are shown in the 
two sub-panels. 
Using the value for the primary mass of $M_\star = 0.96\pm0.04$ $M_\odot$ 
obtained as described in Sect.~3, we derive a minimum
mass for the companion $m_2\sin i = 6.86\pm0.71$ $M_\mathrm{Jup}$. 
The radial-velocity variations measured in HD 81040 thus reveal 
the presence of a long-period ($P\simeq 2.7$ years),
eccentric ($e = 0.53$), massive planetary-mass object. Due to 
gaps in phase coverage and limited number and time baseline of the 
observations, further measurements will be required to improve on 
the determination of some of the orbital elements (such as the 
orbital period and the eccentricity).

Given the values of $m_2\sin i$ and $P$, translating into 
an astrometric signature $\alpha\approx 410$ $\mu$as at the
distance of HD 81040, we have investigated the $Hipparcos$
Intermediate Astrometric Data (IAD) in an attempt to further
constrain the companion mass (for a review of the method 
see for example Sozzetti 2005, and references therein). The star
is flagged as single in the $Hipparcos$ database. The best-fit
astrometric solution with the period of the spectroscopic orbit 
gives a statistically insignificant semi-major axis of $2.0\pm1.0$ mas.
We have then used these values to derive an upper limit to the actual size of the 
astrometric orbit $\alpha$ with a $2.3\sigma$ confidence, following 
the procedure described in Zucker \& Mazeh (2001). 
The resulting upper limit on the companion mass of 
$\sim 75$ $M_\mathrm{Jup}$, at the 99\% confidence level, 
clearly indicates its sub-stellar nature. 

The scatter $\sigma_{O-C}$ of the post-fit residuals reported
in Table~\ref{tab2} is abnormally large when compared to
the average $<\sigma_\mathrm{RV}>$ of the internal errors for both
instruments. Inspection of the velocity residuals (top and bottom 
sub-panels of Fig.~\ref{fig1}) by means of a periodogram search 
revealed no additional periodicity within the time-span of the 
observations. No significant velocity trend can also be found in the 
post-fit residuals (a linear fit to the residuals time series had 
an insignificant slope of $1.8\pm2.2$ m s$^{-1}$ yr$^{-1}$). 
Thus, another companion on a longer-period orbit superposing a
second radial-velocity signal is an unlikely explanation (although
further measurements will help to better address this question).
Another possibility to explain the excess jitter in the residuals
is to invoke effects due to chromospheric activity of the parent
star (e.g., Baliunas et al. 1995; Saar et al. 1998; Santos et al. 2000a).
We discuss this possibility in the next Section.

\section{Properties of the Host Star}

\begin{table}
\caption{Observed and inferred stellar parameters for HD 81040. 
Spectral type, apparent magnitude, colour index, parallax, 
and proper motions are from $Hipparcos$ (ESA 1997). 
The atmospheric parameters $T_\mathrm{eff}$, $\log g$, 
$\xi_t$, and [Fe/H], and the stellar mass and radius 
have been derived as described in Sect.~3.1. The values 
of $v\sin i$, $<\log R^\prime_\mathrm{HK}>$, $\log\epsilon$(Li), 
and age have been obtained as described in Sect.~3.2.}
\label{tab3}
\centering
\begin{tabular}{lc}
\hline\hline
Parameter & Value \\
\hline
$Sp.\, Type$     &   G2/G3 \\
$m_V$   &   $7.72$ \\
$B-V$      &  $0.68$ \\
$\pi$ (mas)     & $30.71\pm 1.24$  \\
$d$ (pc)     & $32.56\pm1.31$  \\
$\mu_\alpha$ (mas yr$^{-1}$)     &  $-151.35\pm 1.08$ \\
$\mu_\delta$ (mas yr$^{-1}$)     &  $35.91\pm 0.52$ \\
$M_V$     & $5.17$  \\
$B.C.$    &  $0.10$ \\
$(U,V,W)$ (km s$^{-1}$)   &  (40.9,$-1.5$,25.1) \\
$T\mathrm{eff}$ (K)     &  $5700\pm 50$ \\
$\log g$ (cgs)     &  $4.5\pm 0.1$ \\
$\xi_t$ (km s$^{-1}$)    &  $0.95\pm 0.05$ \\
$[\mathrm{Fe/H}]$    &  $-0.16\pm 0.06$ \\
$M_\star$ (M$_\odot$)    &  $0.96\pm0.04$ \\
$R_\star$ (R$_\odot$)    &  $0.86\pm 0.04$ \\
$<\log R^\prime_\mathrm{HK}>$        &  $-4.48$    \\
$v\sin i$     &  $2\pm 1$ \\
$\log\epsilon$(Li) &  $1.90/1.91\pm0.07$ \\
$<t>$ (Gyr)    &  $0.73\pm0.1$ \\
\hline
\end{tabular}
\end{table}

HD 81040 (HIP 46076, BD+20\degr2314) is a bright, nearby dwarf. 
$Hipparcos$ astrometry places it at 32.56 pc from the Sun.
The quoted visual magnitude and color index from $Hipparcos$
are $m_V = 7.72$ and $B-V = 0.68$, respectively. The corresponding
absolute magnitude is $M_V = 5.17$, consistent with a
G2/G3 spectral type (Cox 2000), and this classification is 
also favored by our effective temperature determination (see Sect. 3.1). 
A G0 spectral type is reported 
in SIMBAD and in several works in the literature, however the ultimate 
source is the Henry Draper catalog, in which the spectral 
classification is rather coarse. 
Using proper motions and parallax from $Hipparcos$ and the $\gamma$
velocity obtained from the orbital solution described in Sect.~2
(in good agreement with the Nordstr\"om et al. (2004) reported
value of $48.9\pm 0.2$ km s$^{-1}$), the Galactic velocity
vector for HD 81040 (corrected for the Local Standard of Rest following 
Mihalas \& Binney 1981) 
is then $(U,V,W) = (40.9,-1.5,25.1)$ km s$^{-1}$.  These
values are summarized in Table~\ref{tab3}.

\subsection{Abundance Analysis}

We have utilized the Keck/HIRES high-signal-to-noise template
spectrum of HD 81040 to carry out an LTE spectroscopic iron abundance analysis.
In this study, we have followed the approach described in previous works
(Sozzetti et al. 2004, 2005b, and references therein). Our
final iron list consisted of 28 Fe I and 6 Fe II lines
(taken from Sozzetti et al. 2005b). The final
atmospheric parameters, summarized in Table~\ref{tab3}, are:
$T_\mathrm{eff} = 5700\pm 50$ K, $\log g = 4.5\pm 0.1$ (cgs), 
$\xi_t = 0.95\pm 0.05$ km s$^{-1}$, 
and [Fe/H] = $-0.16\pm 0.06$. All these numbers are in very good
agreement with those reported by Allende Prieto \& Lambert (1999)
and Nordstr\"om et al. (2004).

The spectroscopically determined values of $T_\mathrm{eff}$
and [Fe/H], and the absolute luminosity estimate,
were then used as input to the Yale stellar evolution models
(Yi et al. 2003) to derive estimates of the stellar mass and radius for
HD 81040, as well as their uncertainties. The results, also reported
in Table~\ref{tab3}, are as follows: $M_\star = 0.96\pm 0.04$ $M_\odot$,
$R_\star = 0.86\pm 0.04$ $R_\odot$. The predicted $\log g$ value is 4.52, in
excellent agreement with the spectroscopic estimate.

\subsection{Youth and Activity Indicators}

\begin{figure}[t]
\centering
\includegraphics[width=1.\columnwidth]{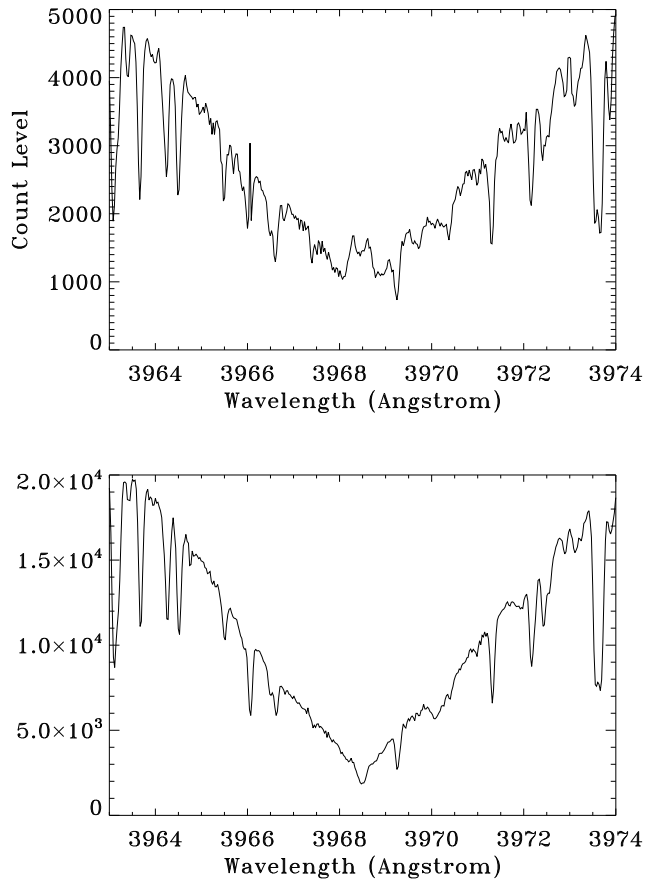}
\caption{Comparison between the Ca II H line for HD 81040
(upper panel) and an inactive star of the same temperature (lower
panel).} \label{activity1}
\end{figure}

Montes et al. (2001), on the basis of its galactic kinematics,
classify HD 81040 as a young Galactic disc star, with no clear membership
to any stellar kinematic group. The star had been selected in that
study based on its chromospheric activity levels as measured by
Strassmeier et al. (2000) in the Vienna-KPNO search for
Doppler-imaging candidate stars. For HD 81040, Strassmeier et al.
(2000) measured a value of the chromospheric emission ratio
$\log R^\prime_\mathrm{HK}= -4.52$.
All our Keck/HIRES spectra show significant core reversal
of the Ca II H and K lines. In the top panel
of Fig.~\ref{activity1} we show a region of the
HIRES template spectrum centered on the Ca II H line.
The emission feature is clearly visible. For
comparison, in the bottom panel we show the spectrum of an old,
inactive star (HIP 105888) with the same temperature from the Sozzetti et
al. (2005a) sample of metal-poor stars, which has a metallicity of
[Fe/H] $= -0.72$.

We have measured the Mount Wilson 
chromospheric activity index S (Duncan et al. 1991) from the Ca II
H and K lines in the Keck/HIRES spectra, and converted it to
$R^\prime_\mathrm{HK}$, following the approach described in
Sozzetti et al. (2004). Based on four data-points, the average
activity level measured in HD 81040, $<\log R^\prime_\mathrm{HK}>
= -4.48$, agrees very well with that reported by Strassmeier et
al. (2000). Based on the Noyes et al. (1984) empirical
calibrations, the mean inferred chromospheric age for HD 81040 is
then $<t> = 0.73\pm0.1$ Gyr, and the average stellar rotation period
$<P_\mathrm{rot}> = 9.8$ days. On the other hand, the stellar
projected rotational velocity appears low. Its measured value,
using the mean ELODIE cross-correlation function (CCF) dip width, 
is $v\sin i = 2\pm1$ km s$^{-1}$, in excellent agreement with 
the one (2.0 km s$^{-1}$) reported by Nordstr\"om et al. (2004). 
Our estimate of $v\sin i$ is also compatible with the two values 
of 3.7 and 4.9 km s$^{-1}$ (with typical uncertainties of 
$2-4$ km s$^{-1}$) quoted by Strassmeier et al. (2000).

\begin{figure}[!t]
\centering
\includegraphics[width=1.\columnwidth]{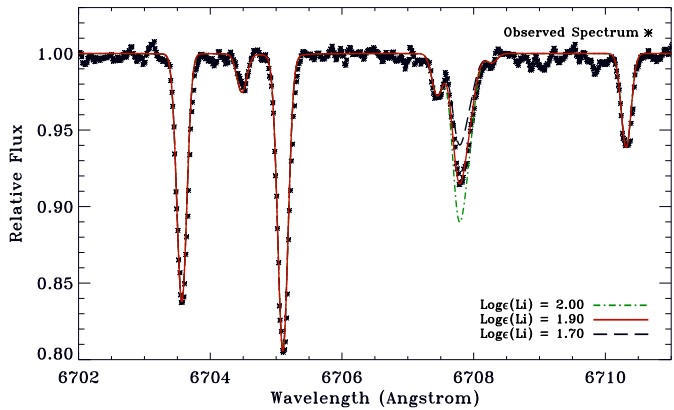}
\caption{A portion of the ELODIE co-added spectrum of
HD 81040 containing the Li I line at 6707.8 \AA\,
(asterisks), compared to three syntheses (lines of different
colours and styles). A significant feature is clearly detected.
\label{activity3}}
\end{figure}

\begin{figure*}[t]
\centering
$\begin{array}{cc}
\includegraphics[width=.45\textwidth]{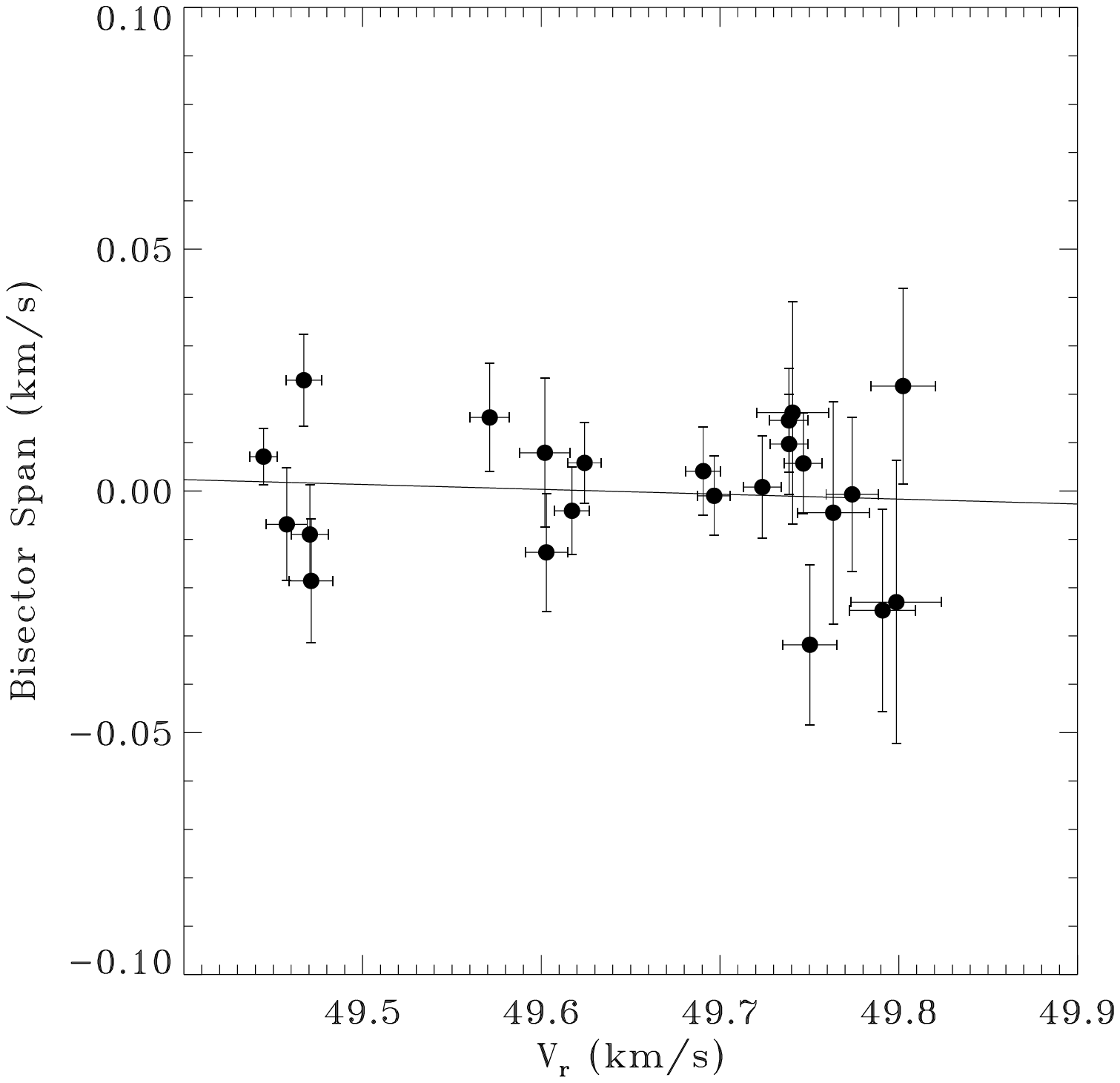} & 
\includegraphics[width=.45\textwidth]{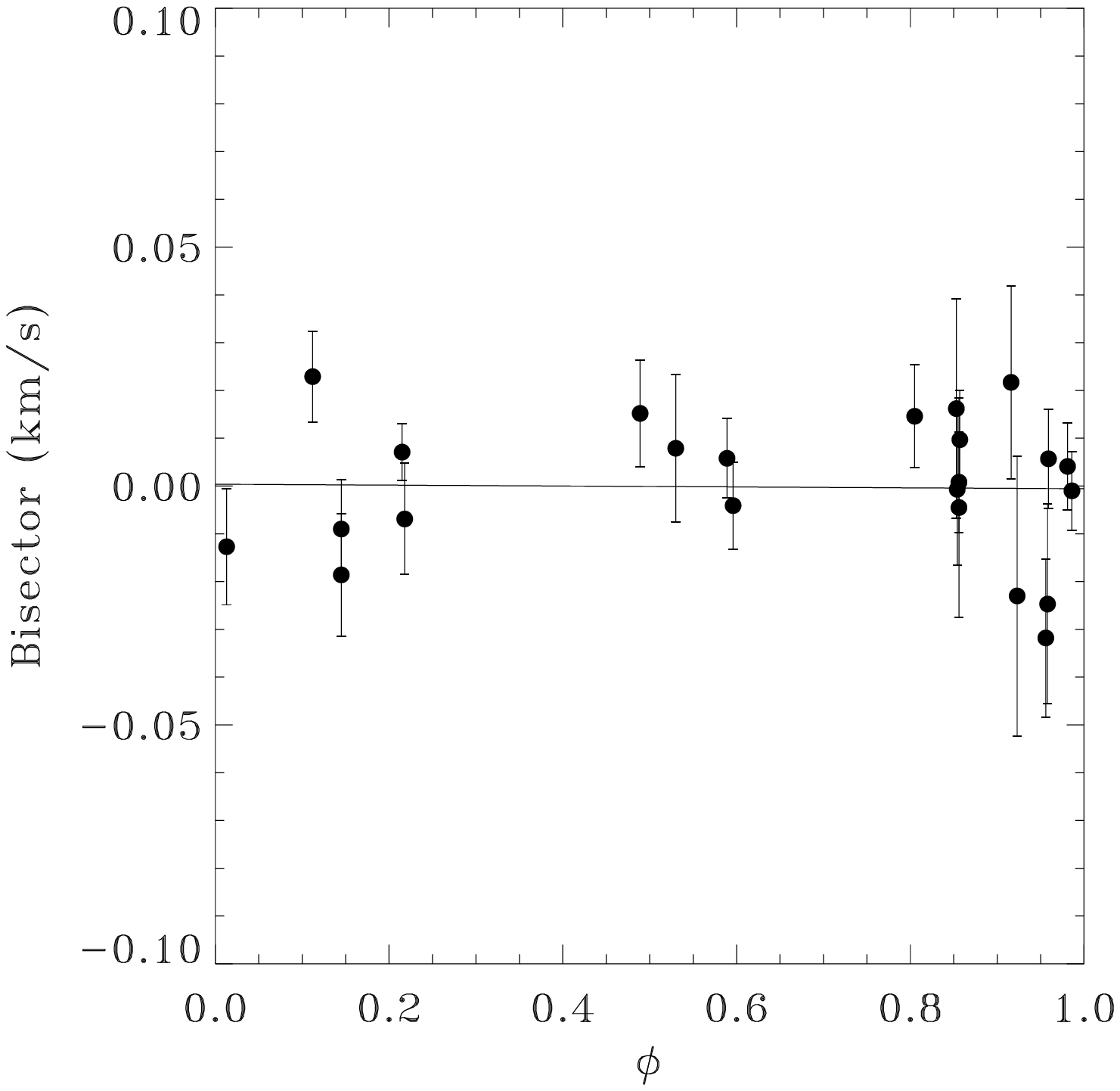} \\
\end{array} $
\caption{Left: Bisector inverse slope of the 
ELODIE CCFs (as defined in Queloz et al. 2001) as a function of the
radial velocities obtained from the CCFs. A linear fit to
the data (solid line) has a statistically insignificant
slope of $-0.010(\pm0.026)V_r$. A rank-correlation 
test gave a probability of no-correlation $p = 0.41$. 
Right: Bisector span as a function of orbital phase 
$\phi$. A linear fit to the data (solid line) has a 
statistically insignificant slope of $-0.0009(\pm009)\phi$. 
A rank-correlation test gave a probability of 
no-correlation $p = 0.58$.} \label{activity4}
\end{figure*}

Another important piece of
circumstantial evidence in favor of a young age for HD 81040 is
the presence of a significant Lithium (Li) feature. In order to
derive an estimate of the Li abundance, we summed all the ELODIE
spectra in the region of the $\lambda = 6707.8$ \AA\, line. We
then utilized the resulting co-added ELODIE spectrum to carry out
a spectral synthesis of the Li line, using the atmospheric
parameters derived from the Fe-line analysis and the same line
list of Reddy et al. (2002). In Fig.~\ref{activity3} we show the
comparison of the spectrum of HD 81040 with three different
models, differing only in the Li abundance. The best-fit model
gives as a result $\log\epsilon$(Li) = 1.90. Following 
Naef et al. (2001b) and Perrier et al. (2003), we also 
directly measured the equivalent width EW$_\lambda$ of the 
$\lambda = 6707.8$ \AA\, Li I line, and used the 
Soderblom et al. (1993) curves of growth to obtain a second, 
independent assessment of the Li abundance. The resulting value 
is $\log\epsilon$(Li) $= 1.91\pm0.07$, with the reported 
error having been estimated by changing EW$_\lambda$ and $T_\mathrm{eff}$ 
by $\pm1\sigma$. Thus, the two methods we applied for the determination 
of the Li abundance in HD 81040 are in very good agreement with each other, 
and our numbers are also compatible with the value of 
$\log\epsilon$(Li) $= 2.13\pm0.15$ quoted by 
Strassmeier et al. (2000). The estimated Li 
abundance is consistent with the typical values for a star 
slightly older than the Hyades cluster (in good agreement 
with the chromospheric age estimate) and with the same 
temperature of HD 81040, both on an empirical as well as 
theoretical basis (see for example Sestito and Randich 2005). 

\subsection{Study of Radial-Velocity Jitter}

With all the information gathered on the parent star pointing
in the direction of a young age for HD 81040, the observed extra-scatter
$\sigma^\prime_v=\sqrt{\sigma^2_{O-C}-<\sigma_\mathrm{RV}>^2}\approx 22$
m s$^{-1}$ in the velocity residuals can then be 
interpreted in terms of activity-related processes. It is well
known (e.g., Saar et al. 1998; Santos et al. 2000a) that
spectral line profile variations induced by surface activity
(e.g., spots) can translate into excess radial-velocity jitter.
For an early-G dwarf with the level of activity exhibited
by HD 81040, empirical estimates 
(e.g., Saar et al. 1998; Santos et al. 2000a; Paulson et al. 2002; 
Wright 2005) indicate a typical value of the activity-induced
radial-velocity jitter of $\sigma^\prime_v\simeq 20$ m s$^{-1}$,
in good agreement with the observed scatter in the post-fit
velocity residuals for HD 81040.

However, the magnitude and period of the detected Doppler
signature are such that its interpretation in terms of the
presence of a massive planetary companion is convincing. In
fact, the estimated rotation period is about two orders of
magnitude shorter than the inferred orbital period, and the
radial-velocity semi-amplitude is almost an order of magnitude
larger than the presumed activity-induced level of jitter. Unlike
the case, for example, of the short-period variable HD 166435
(Queloz et al. 2001), stellar surface activity alone is thus a
very improbable cause for the observed radial-velocity curve. To
put this conclusion on firmer grounds, we have carried out an
analysis of the line bisectors from the ELODIE CCFs, searching for
possible correlations between the bisector span and
radial-velocity measurements, orbital phase, and velocity
residuals. Rank-correlation tests gave probabilities of no
correlation ranging between 0.41 and 0.92. As an example, in 
the two panels of Fig.~\ref{activity4} we show the bisector span 
values plotted against the velocities obtained from the ELODIE CCFs 
and the orbital phase, respectively. No trend is 
visible, further supporting the explanation of true stellar reflex
motion induced by an orbiting companion as opposed to radial-velocity
variations originating in the stellar atmosphere. Unfortunately, 
the $R^\prime_\mathrm{HK}$ measurements are too few to undertake 
a similar analysis. Finally, 
$Hipparcos$ lists HD 81040 as photometrically stable, with a
scatter of 9 mmag. No significant signal at any frequency was
found as a results of a periodogram analysis on the photometric
data. 

\section{Summary and Discussion}

\begin{figure}[t]
\centering
\includegraphics[width=1.\columnwidth]{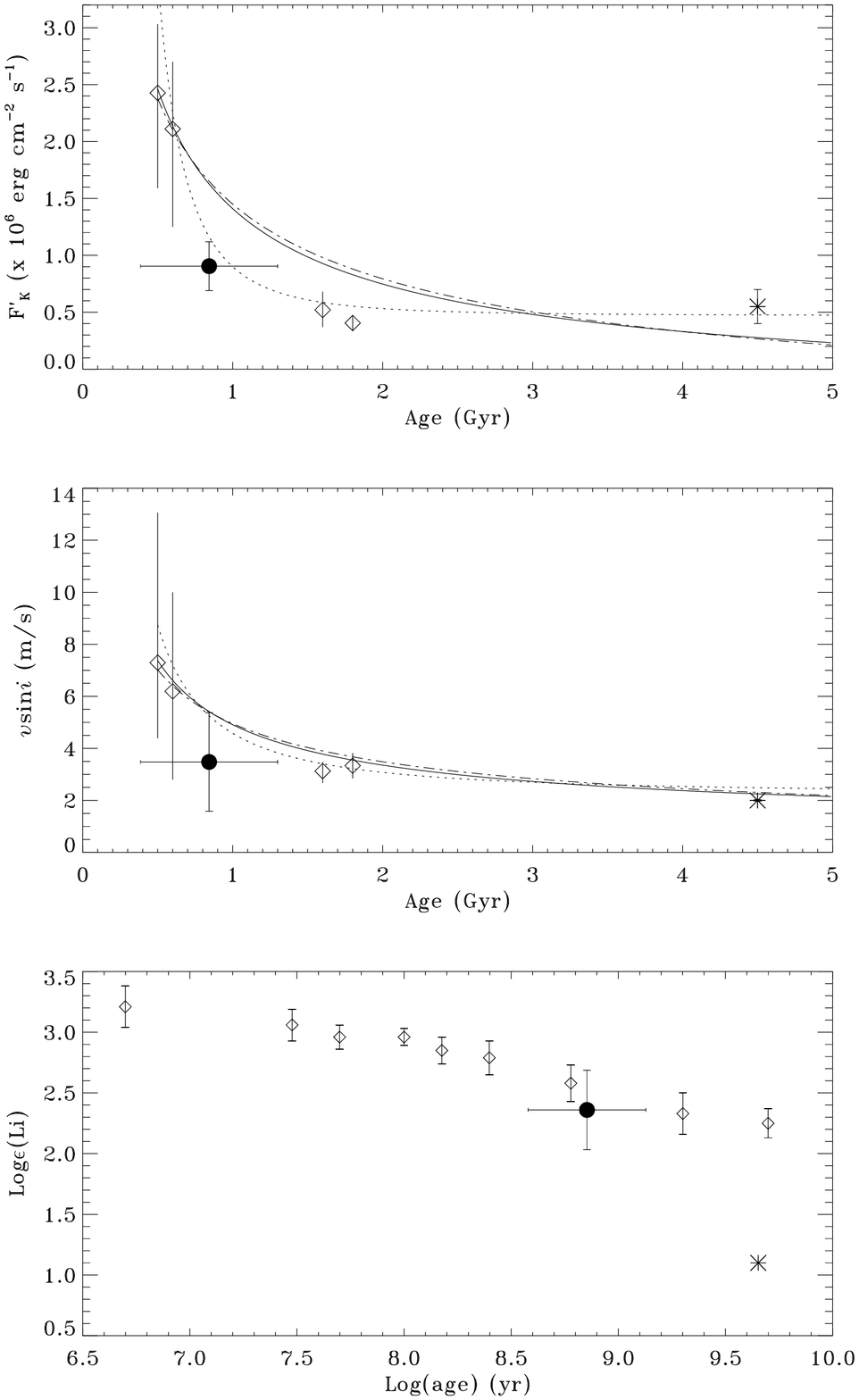}
\caption{Top: average chromospheric flux $F^\prime_\mathrm{K}$ 
(corrected for the photospheric-flux contribution) as a 
function of age for four intermediate-age clusters (diamonds), the sun (asterisk),
and active stars with planets (filled circle). For clusters, error bars
represent peak-to-peak spreads. For stars with planets, they correspond to 
the spread around the mean value. Lines of different shapes
represent theoretical power-law fits to the data with different
exponents (Pace \& Pasquini 2004). Center: the
same, but in the $v\sin i$-age plane. Bottom: average Lithium abundance 
$\log\epsilon$(Li) as a function of age for open clusters (diamonds) 
and active planet hosts (filled circle). For the cluster averages, 
only G-type stars are considered (Sestito and Randich 2005). 
For stars with planets, only actual $\log\epsilon$(Li) estimates 
(no upper limits) have been used. All error bars correspond to 
the dispersion around the mean value.} \label{activity2}
\end{figure}

\begin{table*}
\begin{minipage}[t]{2.\columnwidth}
\caption{Main properties of active planet-hosting stars and their
orbiting planets. Only objects with a chromospheric age estimate  $t < 1.5$ Gyr
are included (corresponding to $\log R^\prime_\mathrm{HK} > -4.65$). Columns
1 through 12 report: star name, spectral type, metallicity, projected rotational velocity,
chromospheric emission ratio, lithium abundance or upper limit (where applicable),
estimated rotation period, chromospheric age, radial velocity jitter estimate,
orbital periods of the companions, minimum masses, and radial-velocity
semi-amplitude. Unless otherwise noted, all parameter values are taken from
the literature sources listed in column 12. The star HD 166435 does not have
a planet, but is listed for comparison.}
\label{tab4}     
\centering                         
\renewcommand{\footnoterule}{}  
\begin{tabular}{l c c c c c c c c c c c c}       
\hline\hline                
Star & $Sp.$ & [Fe/H]& $v\sin i$& $\log R^\prime_\mathrm{HK}$ &
$\log\epsilon$(Li)\footnote{All abundances and upper limits are
taken from Israelian et al. (2004), except for HD 81040 (this
work), HD 40979 (Fischer et al. 2002) and HD 166435 (Queloz et al.
2001).} & $P_\mathrm{rot}$\footnote{From the Noyes et al. (1984)
calibrations.} & Age\footnote{From the Noyes et al. (1984)
calibrations.} & $\sigma^\prime_v$\footnote{The two values are
derived using the Saar et al. (1998) and the Santos et al. (2000a)
calibrations, respectively.} & $P$ & $m_2\sin i$ & $K_1$ &
Refs.\footnote{(1): this work; (2): Naef et al. 2001b; (3)
K\"urster et al. (2000); (4): Hatzes et al. 2000; (5): Fischer  et
al. 2002; (6): Udry et al. 2003; (7): Mayor et al. 2004; (8): Vogt
et al. 2005; (9): Udry et al. 2000;
(10): Udry et al. 2002; (11): Santos et al. 2000b; (12): Queloz et al. 2001.}\\
& $Type$ & &(m s$^{-1}$) & & & (days) &(Gyr) & (m s$^{-1}$)&(days) & ($M_\mathrm{Jup}$) &(m s$^{-1}$) &  \\
\hline
HD 81040 &G2/G3 & $-0.16$ & 2.0  & $-4.48$& 1.90& 10   & 0.7  &  7/19 & 1001  & 6.86 & 168 & (1)\\
HD 1237  &G6V & $+0.10$ & 5.5  & $-4.44$\footnote{From Henry et al. (1996). Naef et al.
(2001b) report a substantially higher value $\log R^\prime_\mathrm{HK} = -4.27$ (but with
large scatter), corresponding to $P_\mathrm{rot}=4$ days
and $t = 0.02$ Gyr.}& 2.24 & 10& 0.6 &  21/21 & 133& 3.32 & 164 & (2)\\
HD 17051 &G0V & $+0.03$ & 6.1  &$-4.65$& 2.63& 12 &1.5& 13/14& 320 & 2.26 & 67& (3)\\
HD 22049 &K2V & $-0.10$ & 2.5\footnote{From Tokovinin (1992). Uncertainties on the
$v\sin i $ value for HD 22049 are rather large.}&$-4.47$\footnote{From Henry et al. (1996).} &
$<0.3$ & 10    & 0.7  & 9/9 & 2502  & 0.86 &  18 & (4)\\
HD 40979 &F8 & $+0.19$ & 7.4  &$-4.63$ &  2.79&   12    & 1.0  & 19/15& 263   & 3.28 & 100 & (5)\\
HD 73256 &G8/K0 & $+0.29$ & 3.2  & $-4.49$& \dots& 12  & 0.8 &  13/19 & 250   & 1.87 & 265 & (6)\\
HD 121504&G2V & $+0.16$ & 2.6  &$-4.57$ &  2.66&   9    & 1.1  &  7/16 & 63    & 1.22 & 45 & (7)\\
HD 128311&K0 & $+0.08$ & 5.7  &$-4.44$&$< -0.4$ &10&0.5 & 11/9& 458& 2.18 & 67 & (8)\\
         & &&&&       &   & &      & 928 & 3.21 &  76 & \\
HD 130322&K0 & $-0.02$ & 1.9  & $-4.39$& $<0.2$& 8   & 0.3  & 7/9 & 10    & 1.02 & 115 &  (9)\\
HD 141937&G2/G3V & $+0.01$ & 2.1  &$-4.65$ &  2.48&   13    & 1.5  & 6/14 & 653   & 9.70  & 247 & (10)\\
HD 142415&G1V & $+0.21$ & 3.3  &$-4.55$ &  \dots&   10    & 1.1  & 8/16& 386   & 1.62 & 51 & (7)\\
HD 147513&G3/5V & $+0.06$ & 1.5  &$-4.38$ &  2.05&   5    & 0.3  &   4/23 & 528   & 1.21 & 31 & (7)\\
HD 192263&K2V & $-0.14$ & 1.8  &$-4.39$ & $<-0.3$& 9\footnote{A photometric period of
24 days, matching the orbital period, was obtained by Henry et al. (2002). However,
Santos et al. (2003) have shown the planet hypothesis still holds.}&0.3& 6/9 & 24&0.72&62&(11)\\
HD 196050&G3V & $+0.22$ & 3.1  &$-4.65$\footnote{Henry et al. (1996) report a much lower
value $\log R^\prime_\mathrm{HK} = -5.04$.}& 2.15 &16 & 1.5& 9/14& 1321 & 3.02 & 49& (7)\\
HD 166435&G0 & $-0.07$ & 7.6  &$-4.26$ & $< 1.7$ & 4\footnote{The photometric period
matches the orbital and chromospheric period estimates. Queloz et al. (2001) have shown the
observed signal is due to activity-related processes.}
    & 0.2  & 33/28 & 4 & 0.60 & 83 & (12)\\
\hline                                   
\end{tabular}
\vfill
\end{minipage}
\end{table*}

We have presented in this work a combined Keck/HIRES + ELODIE
dataset of radial-velocity measurements for the young disc star HD 81040.
The measurements reveal the presence of a massive
($m_2\sin i = 6.86$ $M_\mathrm{Jup}$) planetary companion,
on an eccentric ($e = 0.53$), relatively long-period ($P = 1001$ days) orbit.
The host star is a G2/G3 dwarf with a metallicity [Fe/H] $= -0.16$, 
very close to the average ([Fe/H] $\simeq -0.1$) of the solar neighbourhood 
(Nordstr\"om et al. 2004), 
but somewhat metal-deficient with respect to the average ([Fe/H] $\simeq 0.14$)
of the metallicity distribution of planet-hosting stars (e.g., Fischer \& 
Valenti 2005). 
Both the chromospheric activity level (measured using the Ca II H and K lines
in the HIRES spectra) as well as the Lithium abundance
(estimated using the co-added ELODIE spectrum) speak in favor of a young age
for HD 81040 ($\approx 0.8$ Gyr), and our results agree well with previous
studies of this object (Strassmeier et al. 2000; Montes et al. 2001).

The excess scatter in the residuals from the best-fit orbit with
respect to the nominal internal errors of the HIRES and ELODIE
measurements, $\sim 22$ m s$^{-1}$, is likely attributable to the
high level of activity of HD 81040. Radial-velocity surveys of
nearby F-G-K dwarfs tend to avoid chromospherically
active stars, for which intrinsic velocity jitter caused by
surface inhomogeneities (e.g., spots, convection) not only
degrades the achievable measurement precision, but, when
correlated with the stellar rotation period, can even mimic the
signal produced by an orbiting companion (e.g., Queloz et al.
2001). However, HD 81040 is not the first young star to have been found
harboring a planetary-mass companion, despite its significant
activity levels. We summarize in Table~\ref{tab4} the properties
of active ($\log R^\prime_\mathrm{HK} > -4.65$, corresponding to a
chromospheric age estimate $t < 1.5$ Gyr) stars with known
detected planets, and the main characteristics of the latter. From inspection 
of the Table, at least two important considerations can be made. 

First, the two main youth indicators ($\log R^\prime_\mathrm{HK}$ and, 
when measurable, $\log\epsilon$(Li)), and to a lesser extent $v\sin i$, 
are in fair agreement with each other to indicate relatively young ages 
for these stars. We show in the top and middle panels of Fig.~\ref{activity2} a
comparison between the average values of $v\sin i$ and of 
the chromospheric flux of the Ca II K line core 
$F^\prime_\mathrm{K}$ (expressed in erg cm$^{-2}$ s$^{-1}$) 
for the ensemble of active planet hosts in Table~\ref{tab4} 
and the age-activity-rotation 
relationships for open clusters and average observed values 
from Pace \& Pasquini (2004), and references
therein. The agreement is broad, and a similar result holds for 
the comparison between average Li abundances in open clusters 
from Sestito \& Randich (2005) and the average $\log\epsilon$(Li) 
for the ensemble of active stars with planets (lower panel of 
Fig.~\ref{activity2}). The average 
age inferred from the chromospheric activity values for the 
stars in Table~\ref{tab4} is $\sim 0.8$ Gyr, slightly older than the 
Hyades cluster, and indeed the 
average values of $F^\prime_\mathrm{K}$, $v\sin i$, and $\log\epsilon$(Li) 
are all consistent with such interpretation.

Second, to appreciate how severe a danger (moderate) youth and activity 
can pose for radial-velocity planet surveys, it is useful to compare 
the values of radial-velocity semi-amplitude and orbital period of the 
companions with the expected levels of jitter (carrying typical 
uncertainties of order $30\%-50\%$) and 
estimated stellar rotation periods reported in Table~\ref{tab4}. 
When both $K_1>>\sigma^\prime_v$ and $P>>P_\mathrm{rot}$, conditions realized 
in the majority of the cases presented in Table~\ref{tab4}, including the one 
studied here, little doubt can then be cast on the presence of the
planet. If $K_1>\sigma^\prime_v$, but $P$ is similar to
$P_\mathrm{rot}$ (as is the case for HD 130322 and HD 192263),
simultaneous radial-velocity, photometry, Ca II, and line bisector
measurements must be carried out in order to ascertain that the
planet is real and the observed signal is not intrinsic to the
star (as in the case of HD 166435, also reported in
Table~\ref{tab4} for comparison). If instead $P>>P_\mathrm{rot}$,
but $K_1$ is comparable to $\sigma^\prime_v$, 
the situation becomes more complex, as long-term activity
cycles may reproduce very low-amplitude pseudo-periodic signals,
which vary from season to season (e.g., Cumming et al. 1999). 
For example, based on the absence of a 
periodicity in the $R^\prime_\mathrm{HK}$ time-series similar to
that of the Keplerian signal, Hatzes et al. (2000) claimed HD 22049
actually harbors a planet, but this evidence had not been universally 
accepted in the past as a strong argument in favor of the planet hypothesis
(e.g., Butler et al. 2003). In this specific case, 
the planet existence has indeed been proved correct 
(G. F. Benedict 2005, private communication) by the recently 
completed $HST$ astrometry campaign (Benedict et al. 2003, 2004) 
on this star.

Based on all the evidence presented here, however, the existence of a 
companion around HD 81040 appears to be the best explanation 
for the observed radial-velocity variations. The inferred
minimum mass for the object places it in the high-mass tail of the
extrasolar planet distribution, i.e. the 18\% or so of detected
planets around nearby F-G-K stars with $m_2\sin i \gtrsim 5$
$M_\mathrm{Jup}$. This region of the planet mass distribution is
populated by objects with minimum masses uncomfortably close to
the widely used arbitrary dividing line between planets and brown
dwarfs of 13 $M_\mathrm{Jup}$ (e.g., Oppenheimer et al. 2000). 
As a matter of fact, the vanishing high-mass tail of the
planet mass distribution and the existence of a dearth of close-in
($a \lesssim 5$ AU) brown-dwarf ($M \lesssim 80$ $M_\mathrm{Jup}$)
companions to solar-type stars (the so-called ``brown dwarf
desert''; e.g., Halbwachs et al. 2000) pose a number of puzzles to
models of the formation and dynamical evolution of such low-mass
objects, beginning with their actual nomenclature. Indeed, a
comparatively small number of studies has specifically focused on
explaining the existence of systems such as HD 168443 (the primary 
is orbited by two objects with $m\sin i = 7.7$ $M_\mathrm{Jup}$ 
and $m\sin i = 17.2$ $M_\mathrm{Jup}$. Marcy et al. 2001; 
Udry et al. 2002) or HD 202206 (the primary is orbited by two objects with 
$m\sin i = 17.4$ $M_\mathrm{Jup}$ and $m\sin i = 2.4$ $M_\mathrm{Jup}$; 
Correia et al. 2005), as opposed to the large 
body of work devoted to the study of increasingly lower-mass ($M <
5$ $M_\mathrm{Jup}$) objects, both single and in multi-component
systems.

High-mass planets (or low-mass brown dwarfs) such as the one
presented here found orbiting HD 81040, however,
are very interesting in their own right, as their observed properties,
as well as the characteristics of the host stars,
can provide very important constraints to proposed formation models.
As an illustrative example, recent models of giant planet formation
by accretion of a rocky core (e.g., Lissauer 1993; Pollack et al. 1996)
can qualitatively reproduce the observed mass distribution of
extrasolar planets (Alibert et al. 2005; Ida \& Lin 2004, 2005), and
particularly for $M\lesssim 5$ $M_\mathrm{Jup}$, down to the
Neptune-mass (and lower) regime of some recently discovered objects
(McArthur et al. 2004; Butler et al. 2004; Santos et al. 2004;
Rivera et al. 2005). On the other hand, the alternative disc-instability
model (e.g., Boss 2001, 2005; Mayer et al. 2004)
predicts that objects formed by this mechanism should preferentially
populate the high-mass tail ($M\gtrsim 5$ $M_\mathrm{Jup}$) of the planet mass
distribution (Rice et al. 2003; Rafikov 2005). With improved statistics
of massive planetary companions and their properties, as a complement
to more refined theoretical studies of the efficiency of high-mass giant
planet formation (and of their relative frequency with respect to lower-mass 
planets), the actual roles of the two proposed 
formation modes could be better understood.

Finally, given the projected separation of $62$ mas, and given the
relatively large mass and presumably rather young age of the companion,
HD 81040 could be an interesting candidate for observations with future,
direct near- and far-infrared imaging surveys of wide-separation 
giant planets (e.g., Burrows 2005, and references therein). Also, the
magnitude of the inferred astrometric signature might also make
this system an attractive target for high-precision ground-based as well as
space-borne astrometric surveys (e.g., Sozzetti 2005, and references therein) 
which will come online in the near future.

\begin{acknowledgements}

A.S. gratefully acknowledges financial support through the
Keck PI Data Analysis Fund (JPL 1262605). S.Z. is grateful for partial 
support from the Jacob and Riva Damm Foundation. G.T. acknowledges partial 
support for this work from NASA Origins grant NNG04LG89G.
We wish to thank D. Yong for valuable discussion. 
The referee, William Cochran, provided very helpful comments 
and suggestions. 
We are grateful to the Observatoire de Haute-Provence for the
generous time allocation.
This research has made use of NASA's Astrophysics Data System Abstract
Service and of the SIMBAD database, operated at CDS, Strasbourg, France.

\end{acknowledgements}


\begin{thebibliography}{}

\bibitem[Alibert et al., 2005]{alibert05}
Alibert, Y., Mordasini, C., Benz, W., \& Winisdoerffer, C. 2005,
\aap, 434, 343
\bibitem[Allende Prieto \& Lambert, 1999]{allende99}
Allende Prieto, C., \& Lambert, D. L. 1999, \aap, 352, 555
\bibitem[Baliunas et al. 1995]{baliunas95}
Baliunas, S. L., et al. 1995, \apj, 438, 269
\bibitem[Baranne et al. 1996]{baranne96}
Baranne, A., et al. 1996, \aaps, 119, 373
\bibitem[Benedict et al., 2003]{benedict03}
Benedict, G. F., et al. 2003, \baas, 35, \#67.05
\bibitem[Benedict et al., 2004]{benedict04}
Benedict, G. F., et al. 2004, \baas, 36, \#42.02
\bibitem[Boss, 2001]{boss01}
Boss, A. P. 2001, \nat, 409, 462
\bibitem[Boss, 2005]{boss05}
Boss, A. P. 2005, \apj, 629, 535
\bibitem[Burrows 2005]{burrows05}
Burrows, A. 2005, \nat, 433, 261
\bibitem[Butler et al., 1996]{butler96}
Butler, R. P., Marcy, G. W., Williams, E., McCarthy, C.,
Dosanjh, P., \& Vogt, S. S. 1996, \pasp, 108, 500
\bibitem[Butler et al., 2003]{butler03}
Butler, R. P., Marcy, G. W., Vogt, S. S., Fischer, D. A., Henry, G. W.,
Laughlin, G., \& Wright, J. T. 2003, \apj, 582, 455
\bibitem[Butler et al., 2004]{butler04}
Butler, R. P., Vogt, S. S., Marcy, G. W., Fischer, D. A.,
Wright, J. T., Henry, G. W., Laughlin, G., \& Lissauer, J. J. 2004,
\apj, 617, 580
\bibitem[Cochran et al., 2002]{cochran02}
Cochran, W. D., Hatzes, A. P., \& Paulson, D. B. 2002, \aj, 124, 565
\bibitem[Correia et al. 2005]{correia05}
Correia, A. C. M., Udry, S., Mayor, M., Laskar, J., Naef, D.,
Pepe, F., Queloz, D., \& Santos, N. C. 2005, \aap, 440, 751 
\bibitem[Cox 2000]{cox00}
Cox, A. N. 2000, Allen's Astrophysical Quantities - 4$^{\mathrm{th}}$ ed.,
New York: Springer-Verlag
\bibitem[Cumming et al. 1999]{cumming99}
Cumming, A., Marcy, G. W., \& Butler, R. P. 1999, \apj, 526, 890 
\bibitem[Duncan et al. 1991]{duncan91}
Duncan, D. K., et al. 1991, \apjs, 76, 383
\bibitem[Fischer et al. 2002]{fischer02}
Fischer, D. A., Marcy, G. W., Butler, R. P., Vogt, S. S., Henry, G. W.,
Pourbaix, D., Walp, B., Misch, A., \& Wright, J. T. 2002, \apj, 586, 1394
\bibitem[Fischer \& Valenti, 2005]{fischer05}
Fischer, D. A., \& Valenti, J. 2005, \apj, 622, 1102
\bibitem[Halbwachs et al. 2000]{halbwachs00}
Halbwachs, J. L., Arenou, F., Mayor, M., Udry, S., \& Queloz, D. 2000,
\aap, 355, 581
\bibitem[Hatzes et al. 2000]{hatzes00}
Hatzes, A. P., et al. 2000, \apj, 544, L145
\bibitem[Henry et al. 1996]{henry96}
Henry, T. J., Soderblom, D. R., Donahue, R. A., \& Baliunas, S. L. 1996,
\aj, 111, 439
\bibitem[Henry et al. 2002]{henry02}
Henry, G. W., Donahue, R. A., \& Baliunas, S. L. 2002, \apjl, 577, L111
\bibitem[K\"urster et al. 2000]{kurster00}
K\"urster, M., Endl, M., Els, S., Hatzes, A. P., Cochran, W. D.,
D\"obereiner, S., \& Dennerl, K. 2000, \aap, 353, L33
\bibitem[Ida \& Lin, 2004a]{ida04}
Ida, S., \& Lin, D. N. C. 2004, \apj, 604, 388
\bibitem[Ida \& Lin, 2005]{ida05}
Ida, S., \& Lin, D. N. C. 2005, \apj, 626, 1045
\bibitem[Israelian et al. 2004]{israelian04}
Israelian, G., Santos, N. C., Mayor, M., \& Rebolo, R. 2004, \aap, 414, 601
\bibitem[Latham 2000]{latham00}
Latham, D. W. 2000, in Disks, Planetesimals, and Planets, F.
Garz\`on, C. Eiroa, D. de Winter, and T. J. Mahoney eds, ASP Conf.
Ser., 219, 596
\bibitem[Lissauer 1993]{lissa93}
Lissauer, J. J. 1993, \araa, 31, 129
\bibitem[Marcy \& Butler, 1992]{marcy92}
Marcy, G. W., \& Butler, R. P. 1992, \pasp, 104, 270
\bibitem[Marcy et al. 2001]{marcy01}
Marcy, G. W., et al. 2001, \apj, 555, 418
\bibitem[Marcy et al. 2005a]{marcy05a}
Marcy, G. W., Butler, R. P., Vogt, S. S., Fischer, D. A., Henry, G. W.,
Laughlin, G., Wright, J. T., \& Johnson, J. A. 2005a, \apj, 619, 570
\bibitem[Marcy et al. 2005b]{marcy05b}
Marcy, G. W., Butler, R. P., Fischer, D. A., Vogt, S. S., Wright, J. T.,
Tinney, C. G., \& Jones, H. R. A 2005b, Progress of Theoretical Physics Supplement, 
158, 24
\bibitem[Mayer et al., 2004]{mayer04}
Mayer, L., Quinn, T., Wadsley, J., \& Stadel, J. 2004, \apj, 609, 1045
\bibitem[Mayor \& Queloz 1995]{mayor95}
Mayor, M., \& Queloz, D. 1995, \nat, 378, 355
\bibitem[Mayor et al. 2004]{mayor04}
Mayor, M., Udry, S., Naef, D., Pepe, F., Queloz, D., Santos, N. C., \&
Burnet, M. 2004, \aap, 415, 391
\bibitem[Mazeh et al., 2000]{mazeh00}
Mazeh, T., et al. 2000, \apj, 532, L55
\bibitem[McArthur et al. 2004]{mcarthur04}
McArthur, B. E., et al. 2004, \apjl, 614, L81
\bibitem[Mihalas \& Binney, 1981]{mihalas81}
Mihalas, D., \& Binney, J. 1981, Galactic astronomy: Structure and kinematics - 
2$^\mathrm{nd}$ edition (San Francisco, CA, W. H. Freeman and Co.)
\bibitem[Montes et al., 2001]{montes01}
Montes, D., L\'opez-Santiago, J., G\'alvez, M. C., Fern\'andez-Figueroa, M. J.,
De Castro, E., \& Cornide, M. 2001, \mnras, 328, 45
\bibitem[Naef et al. 2001a]{naef01a}
Naef, D., et al. 2001a, \aap, 375, L27
\bibitem[Naef et al. 2001b]{naef01b}
Naef, D., Mayor, M., Pepe, F., Queloz, D., Santos, N. C., Udry, S.,
\& Burnet, M. 2001b, \aap, 375, 205
\bibitem[Nordstr\"om et al. 2004]{nordstrom04}
Nordstr\"om, B., et al. 2004, A\&A, 418, 989
\bibitem[Noyes et al. 1984]{noyes84}
Noyes, R. W., Hartmann, L. W., Baliunas, S. L., Duncan, D. K., \& 
Vaughan, A. H. 1984, \apj, 279, 763
\bibitem[Oppenheimer et al. 2000]{oppen00}
Oppenheimer, B. R., Kulkarni, S. R., \& Stauffer, J. R. 2000, in
Protostars and Planets IV, ed V. Mannings, A. P. Boss \& S. S. Russell
(Tucson: University of Arizona Press), 1313
\bibitem[Pace \& Pasquini 2004]{pace04}
Pace, G., \& Pasquini, L. 2004, \aap, 426, 1021
\bibitem[Paulson et al. 2002]{paulson02}
Paulson, D. B., Saar, S. H. Cochran, W. D., \& Hatzes, A. P. 2002,
\aj, 124, 572
\bibitem[Paulson et al. 2003]{paulson03}
Paulson, D. B., Sneden, C., \& Cochran, W. D. 2003, \aj, 125, 3185
\bibitem[Perrier et al., 2003]{perrier03}
Perrier, C., Sivan, J.-P., Naef, D., Beuzit, J. L.,
Mayor, M., Queloz, D., \& Udry, S. 2003, \aap, 410, 1039
\bibitem[Pollack et al. 1996]{pollack96}
Pollack, J. B., Hubickyj, O., Bodenheimer, P., Lissauer, J. J.,
Podolack, M., \& Greenzweig, Y. 1996, \icarus, 124, 62
\bibitem[Queloz et al., 2001]{queloz01}
Queloz, D., et al. 2001, \aap, 379, 279
\bibitem[Rafikov, 2005]{rafikov05}
Rafikov, R. R. 2005, \apjl, 621, L69
\bibitem[Reddy et al. 2002]{reddy02}
Reddy, B. E., Lambert, D. L., Laws, C., Gonzalez, G., \&
Covey, K. 2002, \mnras, 335, 1005
\bibitem[Rice et al., 2003b]{rice03b}
Rice, W. K. M., Armitage, P. J., Bonnell, I. A., Bate, M. R., Jeffers, S. V.,
\& Vine, S. G. 2003, \mnras, 346, L36
\bibitem[Rivera et al. 2005]{rivera05}
Rivera, E., Lissauer, J. J., Butler, R. P., Marcy, G. W.,
Vogt, S. S., Fischer, D. A., Brown, T., Laughlin, G., and 
Henry, G. W. 2005, \apj, 634, 625
\bibitem[Saar et al., 1998]{saar98}
Saar, S. H., Butler, R. P., \& Marcy, G. W. 1998, \apj, 498, L153
\bibitem[Santos et al. 2000a]{santos00a}
Santos, N. C., Mayor, M., Naef, D., Pepe, F., Queloz, D., Udry, S.,
\& Blecha, A. 2000a, \aap, 361, 265
\bibitem[Santos et al. 2000b]{santos00b}
Santos, N. C., Mayor, M., Naef, D., Pepe, F., Queloz, D., Udry, S.,
Burnet, M., \& Revaz, Y. 2000b, \aap, 356, 599
\bibitem[Santos et al. 2004]{santos04}
Santos, N. C., et al. 2004, \aap, 426, L19
\bibitem[Sestito \& Randich, 2005]{sestito05}
Sestito, P., \& Randich, S. 2005, \aap, 442, 615
\bibitem[Soderblom et al. 1993]{soderblom93}
Soderblom, D. R., Jones, B. F., Balachandran, S., Stauffer, J. R., 
Duncan, D. K., Fedele, S. B., \& Hudon, J. D. 1993, \aj, 106, 1059
\bibitem[Sozzetti et al., 2004]{sozzetti04}
Sozzetti, A., Yong, D., Torres, G., Charbonneau, D., Latham, D. W.,
Allende Prieto, C., Brown, T. M., Carney, B. W., \& Laird, J. B. 2004,
\apjl, 616, L167
\bibitem[Sozzetti et al., 2005a]{sozzetti05a}
Sozzetti, A., Latham, D. W., Torres, G., Stefanik, R. P., Boss, A.
P., Carney, B. W., \& Laird, J. B. 2005a, in Gaia: The Three-Dimensional
Universe, ESA-SP, 576, 309
\bibitem[Sozzetti et al., 2005b]{sozzetti05b}
Sozzetti, A., Yong, D., Carney, B. W., Laird, J. B.,
Latham, D. W., \& Torres, G. 2005b, \aj, submitted
\bibitem[Sozzetti, 2005]{sozz05}
Sozzetti, A. 2005, \pasp, 117, 1021
\bibitem[Strassmeier et al., 2000]{strass00}
Strassmeier, K., Washuettl, A., Granzer, Th., Scheck, M., \&
Weber, M. 2000, \aap, 142, 275
\bibitem[Tokovinin 1992]{tokovinin92}
Tokovinin, A. A. 1992, \aap, 256, 121
\bibitem[Udry et al. 2000]{udry00}
Udry, S., Mayor, M., Naef, D., Pepe, F., Santos, N. C, Queloz, D.,
Burnet, M., Confino, B. \& Melo, C. 2000, \aap, 356, 590
\bibitem[Udry et al. 2002]{udry02}
Udry, S., Mayor, M., Naef, D., Pepe, F., Queloz, D., Santos, N. C,
\& Burnet, M. 2002, \aap, 390, 267
\bibitem[Udry et al. 2003]{udry03}
Udry, S., et al. 2003, \aap, 407, 679
\bibitem[Valenti et al., 1995]{valenti95}
Valenti, J. A., Butler, R. P., \& Marcy, G. W. 1995, \pasp, 107, 966
\bibitem[Vogt et al. 1994]{vogt94}
Vogt, S. S., et al. 1994, in Instrumentation in Astronomy VIII, D.
L. Crawford \& E. R. Craine eds., Proc. SPIE , 2198, 362
\bibitem[Vogt et al. 2005]{vogt05}
Vogt, S. S., Butler, R. P., Marcy, G. W., Fischer, D. A., Henry, G. W.,
Laughlin, G., Wright, J. T., \& Johnson, J. A. 2005, \apj, 632, 638
\bibitem[Wright, 2005]{wright05}
Wright, J. T. 2005, \pasp, 117, 657
\bibitem[Yi et al. 2003]{yi03}
Yi, S., Kim, Y. -C., \& Demarque, P. 2003, \apjs, 144, 259
\bibitem[Zucker \& Mazeh, 2001]{zucker01}
Zucker, S., \& Mazeh, T. 2001, \apj, 562, 549
\bibitem[Zucker et al. 2002]{zucker02}
Zucker, S., et al. 2002, \apj, 568, 363

\end{thebibliography}
\end{document}